\documentclass[11pt,onecolumn,aps,prd,amssymb,nofootinbib]{revtex4-2}

\usepackage{amssymb}
\usepackage{lipsum}
\usepackage{amsmath}
\usepackage{graphicx}

\usepackage[format=hang,font=small,labelfont=bf]{caption}
\usepackage[colorlinks=true,citecolor=blue,linkcolor=red,urlcolor=blue]{hyperref}

\begin{document}

\title{A Statistical Derivation of Bekenstein-Hawking Entropy for Schwarzschild Black Holes}
\author{Naman Kumar}
\affiliation{School of Basic Sciences, Indian Institute of Technology, Bhubaneswar, 752050}
\email{namankumar5954@gmail.com}

\begin{abstract}
    A microscopic derivation of the Bekenstein-Hawking entropy for the Schwarzschild black hole was presented earlier by using a non-trivial phase space. It was argued that the Schwarzschild black hole behaves like a 1D quantum mechanical system. In this paper, we show that if we assume the phase space to obey the holographic principle and take the microscopic particles inside the quantum gravitational system to be ideal bosonic gas, we can derive the Bekenstein-Hawking entropy. The assumption of the phase space to follow the holographic principle such that the Schwarzschild black hole behaves as a 2D system is very much in the spirit of our understanding of black holes than their behavior as a 1D system. However, the argument suggests that the black hole be treated as a system with the equation of state $P=\rho$.
    \paragraph*{Keywords}:
Bekenstein-Hawking Entropy, Statistics of Black Holes, Holographic principle, 2D phase space.

\end{abstract}
\maketitle
\section{Introduction}
It is well known that black holes have thermodynamic behaviour. Hawking\cite{hawking1971gravitational} first showed that the area of black holes is a non-decreasing function of time
\begin{equation}
    \frac{dA}{dt}\geq0
\end{equation}
Bekenstein\cite{PhysRevD.7.2333} argued that this similarity can be attached to the thermodynamic concept of entropy and proposed the equation
\begin{equation}
    S=\eta k_B\frac{A}{l^2_p}
\end{equation}
where $\eta$ was an undetermined dimensionless constant. When Hawking\cite{1975CMaPh..43..199H} derived the temperature of the black hole, the speculations turned robust, and the entropy was obtained as
\begin{equation}
    S=k_B\frac{A}{4l^2_p}
\end{equation}
thereby fixing the dimensional constant $\eta$. This is the famous Bekenstein-Hawking entropy. This result is very interesting as it suggests that the entropy scales as the area rather than the conventional volume. This equation further led to the idea of the holographic principle\cite{hooft1993dimensional,susskind1995world}, which suggests that the information about the volume of space is stored on its boundary. The best realization of the holographic principle is the AdS/CFT correspondence\cite{maldacena1999large}. In\cite{xiao2020microscopic}, a microscopic derivation of the Bekenstein-Hawking entropy for Schwarzschild black hole was presented by assuming a non-trivial phase space. The independent quantized modes were calculated by $g\frac{c^3Vdp}{2\pi G\hbar^2}$ rather than the conventional $\frac{Vd^3\vec{p}}{(2\pi\hbar)^3}$, thereby effectively treating the 3+1 Schwarzschild black hole as a $1D$ quantum mechanical system. For example, in the conventional case, the logarithm of the partition function of a photon gas system is given by\cite{huang2009introduction}
\begin{equation}
\ln\Xi=\frac{2V}{(2\pi\hbar)^3}\int\ln(1-e^{-\beta cp})d^3\Vec{p}=\frac{\pi^2}{45c^3\hbar^3}\frac{V}{\beta^3}
\end{equation}
It was further argued that the black hole be viewed as a system with the equation of state $P=\rho$. This equation of state has also been attached with black hole thermodynamics earlier (see \cite{masoumi2014equation} and references therein). Each point in the phase space represents a state of the system. In the spirit of the holographic principle, we take the phase space to be $2D$ such that the state of the system is essentially contained completely on the $2D$ phase space. The paper is organized as follows. We first review the earlier derivation presented in\cite{xiao2020microscopic}. Then we derive the Bekenstein-Hawking entropy formula using a statistical approach. Finally, we discuss the motivation behind statistical derivation based on the emergent nature of gravity and information limit due to the holographic principle. 
\section{A Brief Review of Earlier Derivation}
The following assumptions were made: First, the particles inside the quantum gravitational system are bosonic and massless. Second, they obey the energy-momentum relation $\epsilon=pc$ with $p=|\vec{p}|$ and third, the calculation of quantized modes is done by $g\frac{c^3Vdp}{2\pi G\hbar^2}$. Next, the Schwarzschild black hole of radius $R$ in 3+1 dimensions is modelled as a system consisting of these particles. The logarithm of the partition function, in this case, is given by
\begin{equation}
    \ln Z=-\frac{gc^3V}{\pi G\hbar^2}\int_{0}^{\infty}\ln(1-e^{-\beta cp})dp=\frac{g\pi c^2V}{6G\hbar^2\beta}
\end{equation}
Then we obtain the expressions for entropy and energy as
\begin{equation}
    E=-\frac{\partial}{\partial\beta}Z=\frac{g\pi k_B^2c^2}{6G\hbar^2}VT^2
\end{equation}

\begin{equation}
    S=k_B(\ln Z+\beta E)=\frac{g\pi k_B^2c^2}{3G\hbar^2}VT\label{entropy1}
\end{equation}
The pressure is given by
\begin{equation}
    P=k_BT\frac{\partial\ln Z}{\partial V}=\frac{g\pi k_B^2c^2}{6G\hbar^2}T^2
\end{equation}
Comparing with $\rho=E/V$ gives the equation of state as $P=\rho$. The Komar mass as the gravitational source corresponds to $(\rho+3P)V$, which gives
\begin{equation}
    M=4E=\frac{2g\pi k_B^2c^2}{3G\hbar^2}VT^2=\frac{2g\pi k_B^2c^2}{3G\hbar^2}\frac{4}{3}\pi R^3T^2
\end{equation}
Taking $M$ to be the mass of the black hole and choosing $g=9$ along with $R=\frac{2GM}{c^2}$ for a Schwarzschild black hole gives the Hawking temperature, $T=\frac{\hbar c^3}{8\pi Gk_BM}$. Substituting into Eq.(\ref{entropy1}), we get the Bekenstein-Hawking entropy.

\section{A Statistical Derivation Based On \\Holographic Principle}
We make the following assumptions about the microscopic particles inside a quantum gravitational system. First, the particles are ideal bosonic gas with mass $m$ such that their partition function follows Bose-Einstein statistics and they obey the energy-momentum relation of an ideal gas which is given by $\epsilon=p^2/2m$ where $p=|\vec p|$. Second, the calculation of independent quantized modes follows the holographic principle such that the states of this system are completely specified by a $2D$ phase space and is evaluated by $g\frac{V}{l_p}\frac{d^2\vec{p}}{(2\pi)^2\hbar^2}$=$gV\sqrt{\frac{c^3}{G\hbar}}\frac{d^2\Vec{p}}{(2\pi\hbar)^2}$, where $g$ a dimensionless constant is introduced to include any other degrees of freedom and the gravitational system is effectively 2D with area $A=\frac{V}{l_p}$. We model the Schwarzschild black hole of radius $R$ in 3+1 dimensions as a system consisting of these particles.
The logarithm of the partition function($Z$) is
\begin{equation}
    \ln Z=-\sqrt{\frac{c^3}{G\hbar}}\frac{2gV}{(2\pi\hbar)^2}\int_{0}^{\infty} \ln(1-e^{-p^2/2mk_BT})d^2\vec p\label{partitionbose}
\end{equation}
It is important to remark that if we consider the black hole to consist of a massive particle (ideal bosonic gas in this case), then 2D phase space follows inevitably, as can be verified from Eq.(\ref{partitionbose}) otherwise, we can not recover the Bekenstein-Hawking entropy. Eq.(\ref{partitionbose}) can be re-written as
\begin{equation}
   \ln Z =-\sqrt{\frac{c^3}{G\hbar}}\frac{2gV}{(2\pi\hbar)^2}\int_{0}^{\infty} 2\pi \ln(1-e^{-p^2/2mk_BT})p\hspace{0.5mm}dp=\sqrt{\frac{c^3}{G\hbar}}\frac{2gV}{(2\pi\hbar)^2}\frac{\pi^3m}{3\beta}
\end{equation}
where $\beta=1/k_BT$ and $V=\frac{4}{3}\pi R^3$. We, therefore, get the expressions for the energy and entropy as
\begin{equation}
    E=-\frac{\partial}{\partial\beta}\ln Z=\sqrt{\frac{c^3}{G\hbar}}\frac{2g\pi^3k_B^2}{3(2\pi\hbar)^2}mVT^2
\end{equation}
\begin{equation}
    S=k_B(\ln Z+\beta E)=\sqrt{\frac{c^3}{G\hbar}}\frac{4g\pi^3k_B^2}{3(2\pi\hbar)^2}mVT\label{entropy}
\end{equation}
The pressure is given by
\begin{equation}
    P=k_BT\frac{\partial \ln Z}{\partial V}=\sqrt{\frac{c^3}{G\hbar}}\frac{2g\pi^3m}{3(2\pi\hbar)^2}k_B^2T^2
\end{equation}
Comparing with $\rho=E/V$ gives the equation of state as $P=\rho$.
Komar mass as the source then gives $M=4E$. Thus, we get
\begin{equation}
   Mc^2=4E=\sqrt{\frac{c^3}{G\hbar}}\frac{8g\pi^3k_B^2}{3(2\pi\hbar)^2}mVT^2\label{mass}
\end{equation}
Taking $gm=\frac{9\hbar}{cl_p}=9m_p$ where $m_p$ is Planck mass, and $V=\frac{4}{3}\pi R^3$ such that $R=\frac{2GM}{c^2}$ for a Schwarzschild black hole, we finally obtain using Eq.(\ref{mass})
\begin{equation}
  T=\frac{\hbar c^3}{8\pi Gk_BM}  
\end{equation}
which is the Hawking temperature. Putting this value in Eq.(\ref{entropy}) we get
\begin{equation}
    S=\frac{k_B\pi R^2c^3}{G\hbar}=\frac{k_BA}{4l_p^2}
\end{equation}
where $l_p^2=G\hbar/c^3$. Thus, we obtained the Bekenstein-Hawking entropy by assuming that the phase space follows the holographic principle. On comparing Eq.(\ref{entropy}) and Eq.(\ref{mass}) we get the relation
\begin{equation}
    2TS=M\label{smarr}
\end{equation}
This is the Smarr formula for the 3+1 dimension Schwarzschild black hole. It is important to emphasize here the importance of Eq.(\ref{smarr}). Deducing this equation in this microscopic derivation is necessary to derive the correct form of the Bekenstein-Hawking Entropy; otherwise, we will end up with the wrong coefficient even if we obtain the correct Hawking temperature. Another thing to note is that the equation of state $P=\rho$ is critical to get the correct Smarr formula, and as discussed earlier, the failure to obtain the correct Smarr formula leads to the wrong coefficient. This suggests that the black hole is to be viewed as a system with the equation of state $w\equiv\frac{P}{\rho}=1$, as already argued in \cite{xiao2020microscopic} and references therein. The mass $m$ of each microscopic particle is then interpreted as $m=m_p$ and the dimensionless constant $g=9$.
\section{Conclusion and Discussion}
In this paper, we successfully derived the Bekenstein-Hawking Entropy of a Schwarzschild black hole by considering it to consist of ideal bosonic gas at the microscopic level with a phase space that obeys the holographic principle. It is remarkable that the correct Bekenstein-Hawking temperature can be exactly derived from such a simplistic picture. Moreover, in\cite{zurek1984black,hooft1998self}, the authors reproduced correct Bekenstein-Hawking entropy by taking $w\equiv\frac{P}{\rho}=1$ although the reasons were not fully understood. In\cite{xiao2020microscopic}, the authors managed to exactly "derive" this result. Our derivation in a different setting strongly supports this idea. The phase space is taken to be two-dimensional such that all the information about the bulk system is contained in it. This essentially means treating the black hole as a two-dimensional system at the microscopic scale. The earlier derivation presented in\cite{xiao2020microscopic} treated black holes as a 1D system where the quantum particles were taken to be equivalent to a photon gas. Both these pictures do not seem to be very realistic. On the other hand, treating the black hole as a 2D system is very much in the spirit of our understanding of black holes. The assumption of quantum particles to be ideal boson gas is much more realistic too. It is also widely believed that gravity and spacetime have emergent origins. Verlinde\cite{verlinde2011origin} argued that gravity is an entropic force arising from the tendency of a microscopic theory to maximize its entropy. This shows that there is a maximum limit on the amount of information in a volume of space. This is also the result of the holographic principle. In this view, it becomes important to microscopically derive the Bekenstein-Hawking entropy. This strongly motivated us to study a statistical derivation based on the holographic principle. This derivation also suggests that the black hole be treated as a system with the equation of state $w\equiv\frac{P}{\rho}=1$. A series of works called holographic cosmology has been studied based on this equation of state\cite{banks2001holographic,banks2005holographic}. We are hopeful that these results will inspire further investigation.
\section*{Conflict of Interest}
The author declares no conflict of interest.
\bibliography{bib}
\bibliographystyle{unsrt}

\end{document}